\documentclass[
    prl,
    reprint,
    aps,
    superscriptaddress,
    nobibnotes,
    nofootinbib
]{revtex4-2}

\usepackage[utf8]{inputenc}
\usepackage[T1]{fontenc}
\usepackage[english]{babel}
\usepackage{amsmath, amssymb, amsfonts, bm}
\usepackage{physics}
\usepackage[colorlinks,citecolor=blue,urlcolor=blue,linkcolor=blue]{hyperref}
\usepackage{graphicx}
\usepackage{float}
\usepackage{tikz}
\usepackage{upgreek}
\usepackage[tight]{units}

\newcommand{\ethr}[0]{E_{\mathrm{thr}}}

\newcommand{\ega}{$\upbeta$/$\upgamma$}
\newcommand{\onepi}{{COSINUS-1$\uppi$}}
\newcommand{\twopi}{{COSINUS-2$\uppi$}}

\newcommand{\gas}{$\upgamma$s}

\newcommand{\textalpha}{$\upalpha$}
\newcommand{\textpi}{$\uppi$}
\newcommand{\textnu}{$\upnu$}
\newcommand{\x}{$\times$}
\newcommand{\oom}[1]{$\mathcal{O}(#1)$}

\definecolor{nicered}{RGB}{208,2,27}

\AtBeginDocument{}

\begin{document}

\title{COSINUS -- a model-independent challenge of the DAMA/LIBRA dark matter claim  with cryogenic NaI detectors operated in a new low-background facility}
\date{\today}

\newcommand{\affi}{\affiliation{Max-Planck-Institut f\"ur Physik, 85748 Garching - Germany}}
\newcommand{\affii}{\affiliation{INFN - Sezione di Roma, 00185 Roma - Italy}}
\newcommand{\affiii}{\affiliation{Gran Sasso Science Institute, 67100 L'Aquila - Italy}}
\newcommand{\affiv}{\affiliation{INFN - Laboratori Nazionali del Gran Sasso, 67100 Assergi - Italy}}
\newcommand{\affv}{\affiliation{Institut f\"ur Hochenergiephysik der \"Osterreichischen Akademie der Wissenschaften, 1010 Wien - Austria}}
\newcommand{\affvi}{\affiliation{Dipartimento di Scienze Fisiche e Chimiche, Università degli Studi dell' Aquila, 67100 L'Aquila - Italy}}
\newcommand{\affvii}{\affiliation{Atominstitut, Technische Universit\"at Wien, 1020 Wien - Austria}}
\newcommand{\affviii}{\affiliation{State Key Laboratory of Functional Crystals and Devices, Shanghai Institute of Ceramics, Chinese Academy of Sciences, 201899 Shanghai - China}}
\newcommand{\affix}{\affiliation{Helsinki Institute of Physics, 00014 University of Helsinki - Finland}}
\newcommand{\affx}{\affiliation{SNOLAB, P3Y 1N2 Lively - Canada}}
\newcommand{\affxi}{\affiliation{CNR-SPIN c/o Dipartimento di Scienze Fisiche e Chimiche, Università degli Studi dell’Aquila, 67100 L’Aquila - Italy}}
\newcommand{\affxii}{\affiliation{Present address: Department of Physics, ETH Zurich, CH-8093 Zurich, Switzerland}}
\newcommand{\affxiii}{\affiliation{Present address: ETH Zurich - PSI Quantum Computing Hub, Paul Scherrer Institute,CH-5232 Villigen, Switzerland}}

\newcommand{\corri}{\email{Corresponding author: florian.reindl@tuwien.ac.at}}
\newcommand{\corrii}{\email{Corresponding author: kschaeff@mpp.mpg.de}}
\newcommand{\corriii}{\email{Corresponding author: philipp.schreiner@tuwien.ac.at}}

\author{G.~Angloher}\affi
\author{M.~R.~Bharadwaj}\affi
\author{A.~B\"ohmer}\affv\affvii
\author{M.~Cababie}\affv\affvii
\author{I.~Colantoni}\affii
\author{I.~Dafinei}\affii\affiii
\author{N.~Di~Marco}\affiii\affiv
\author{C.~Dittmar}\affi
\author{L.~Einfalt}\affv\affvii
\author{F.~Ferella}\affiv
\author{F.~Ferroni}\affii\affiii
\author{S.~Fichtinger}\affv
\author{A.~Filipponi}\affiv\affvi
\author{M.~Friedl}\affv
\author{L.~Gai}\affviii
\author{M.~Gapp}\affi
\author{M.~Heikinheimo}\affix
\author{K.~Heim}\affi
\author{M.~N.~Hughes}\affi
\author{K.~Huitu}\affix
\author{M.~Kellermann}\affi\corriii\affv\affvii
\author{R.~Maji}\affv\affvii
\author{M.~Mancuso}\affi
\author{L.~Pagnanini}\affiii\affiv
\author{F.~Petricca}\affi
\author{S.~Pirro}\affiv
\author{F.~Pr\"obst}\affi
\author{G.~Profeta}\affiv\affvi
\author{A.~Puiu}\affiv
\author{F.~Reindl}\corri\affv\affvii
\author{K.~Sch\"affner}\corrii\affi
\author{J.~Schieck}\affv\affvii
\author{P.~Schreiner}\corriii\affv\affvii
\author{C.~Schwertner}\affv\affvii
\author{K.~Shera}\affi
\author{M.~Stahlberg}\affi
\author{A.~Stendahl}\affix
\author{M.~Stukel}\affiv\affx
\author{C.~Tresca}\affiv\affxi
\author{S.~Yue}\affviii
\author{V.~Zema}\affi\affv
\author{Y.~Zhu}\affviii
\author{N.~Zimmermann}\affix

\collaboration{The COSINUS Collaboration}
\noaffiliation

\begin{abstract}
Low-temperature detectors are a powerful technology for dark matter search, offering excellent energy resolution and low energy thresholds. COSINUS is the only experiment that combines scintillating sodium iodide (NaI) crystals with an additional phonon readout at cryogenic temperatures, using superconducting sensors (remoTES), alongside the conventional scintillation light signal. Via the simultaneous phonon and scintillation light detection, a unique event-by-event particle identification is enabled. This dual-channel approach allows for a model-independent cross-check of the long-standing DAMA/LIBRA signal with a moderate exposure of a few hundred kg\,d, while completely avoiding key systematic uncertainties inherent to scintillation-only NaI-based searches. COSINUS built and commissioned a dedicated low-background cryogenic facility at the LNGS underground laboratories. Data taking with eight NaI detector modules (\onepi~Run1) is planned to begin in late 2025. 
\end{abstract}

\maketitle

\section{Introduction} \label{sec:introduction}
Dark matter (DM) remains one of the big mysteries of present-day physics. The existence of this elusive substance is proven by numerous astrophysical and cosmological observations, which are all based on the gravitational pull exerted by DM. Direct detection experiments aim to decipher the so-far unknown nature of DM by directly observing interactions of DM particles in earth-bound detectors. In the past three decades, direct detection experiments achieved tremendous gains in sensitivity over a wide potential DM particle mass range, using different target materials and experimental techniques~\cite{navas_review_2024}.

None of these experiments claims to observe a DM signal, except for the DAMA/LIBRA (formerly DAMA/NaI) collaboration~\cite{bernabei_first_2018,bernabei_further_2021,bernabei_annual_2025}. DAMA observes a statistically robust annual modulation of the event rate in their detectors where the period and phase of the modulation almost perfectly fit the expectation for the presence of a halo of DM particles in the Milky Way \cite{bernabei_first_2018,bernabei_further_2021}. The contradiction between null results on the one side and a positive signal claim on the other side is challenging to resolve in an unambiguous way. Such a cross-check needs an experiment with a sodium iodide (NaI) target, the same target material as used by DAMA. Any other choice would introduce dependencies on the target material and would consequently not be model-independent. 

Several NaI-based searches are already taking data (ANAIS \cite{amare_performance_2019} and COSINE-100 \cite{adhikari_initial_2018}), or are in the planning/construction phase (SABRE \cite{barberio_sabre_2025,antonello_sabre_2019}). These experiments, just like DAMA, measure the scintillation light emitted by thallium-doped NaI crystals at room temperature as their single signal and they succeeded to scale up their target masses to \oom{\text{100\,kg}}. 
The biggest challenge in the comparison of data from scintillation-light-only (NaI-based) experiments is the substantial systematic uncertainty on their quenching factors at low energies and, thus, on their nuclear recoil energy scales~\cite{xu_scintillation_2015,cintas_measurement_2024}. Furthermore, scintillation-only experiments can perform particle identification using pulse-shape analysis only on a statistical basis, rather than on an event-by-event level, which severely limits their ability to distinguish signal events from background events.

The Cryogenic Observatory for SIgnatures seen in Next-generation Underground Searches (COSINUS) takes a novel and unique path for NaI-based experiments. COSINUS operates pure NaI crystals at millikelvin (mK) temperatures, simultaneously reading two signals: the phonon/heat signal and the scintillation light signal. The phonon signal is not quenched and provides a precise reconstruction of the deposited energy and a low threshold for both electron and nuclear recoils. The ratio of light to phonon signal, called \emph{light yield}, enables a powerful event-by-event particle discrimination, and, thus, COSINUS can perform a quasi-background-free probe of DM-nucleus scattering as the origin of the DAMA modulation signal~\cite{Angloher2016}.

COSINUS was constructed from 2021 to 2024 in the Laboratori Nazionali del Gran Sasso (LNGS) underground laboratories in central Italy. This paper presents the COSINUS detector and the experimental setup in a compact form and gives details and projections for the staged physics program, starting with \onepi~Run1 in 2025. The goal for this run is to collect 100\,kg\,d of exposure to cross-check a standard-scenario-like\footnote{Standard scenario denotes a set of assumptions, mainly elastic DM nucleus scattering and a Maxwellian velocity distribution of the DM particles in the Milky Way~\cite{baxter_recommended_2021}.} interaction as an explanation for the DAMA modulation signal. 
After one to two years of data taking, a detector upgrade is planned to substantially increase the exposure to 1000\,kg\,d in \onepi~Run2. After about two more years, it will settle the question of whether DM-nucleus interactions cause the DAMA modulation signal in a fully model-independent and unambiguous way.

\section{The COSINUS detection principle} \label{sec:detector}
Particle interactions in NaI crystals produce two kinds of signals: lattice vibrations (phonons/heat) and scintillation light. The phonons are readily read out from the NaI crystal using a Transition Edge Sensor (TES) capable of measuring the induced temperature rise. The scintillation photons are captured by a beaker-shaped silicon (Si) light detector (and subsequently read out with a separate TES). Such cryogenic solid-state detectors are a well-proven technology for studying rare event processes due to their low-energy thresholds as low as ${\mathcal{O}}$(\unit[1]{eV}) and excellent energy resolutions, and the thin-film and lithography processes required for tungsten (W) TES structures of $\mathcal{O}$(\unit[0.1-1]{\textmu m}) thickness have been well-established by the CRESST dark matter search group at the Max-Planck Institute for Physics in Garching, Germany \cite{probst_model_1995,ferger_cryogenic_1996}.\par In CRESST, the TES is directly fabricated onto the target crystal, which withstands the conditions of thin-film production, wet chemistry, and several iterations of photo-lithography. While this approach also works for the COSINUS light detector, the main challenge was to adapt the procedure to the restrictions imposed by NaI as a target material.

\subsection{Novel remoTES design}
\begin{figure}[t]
    \centering
    \includegraphics[width=\linewidth]{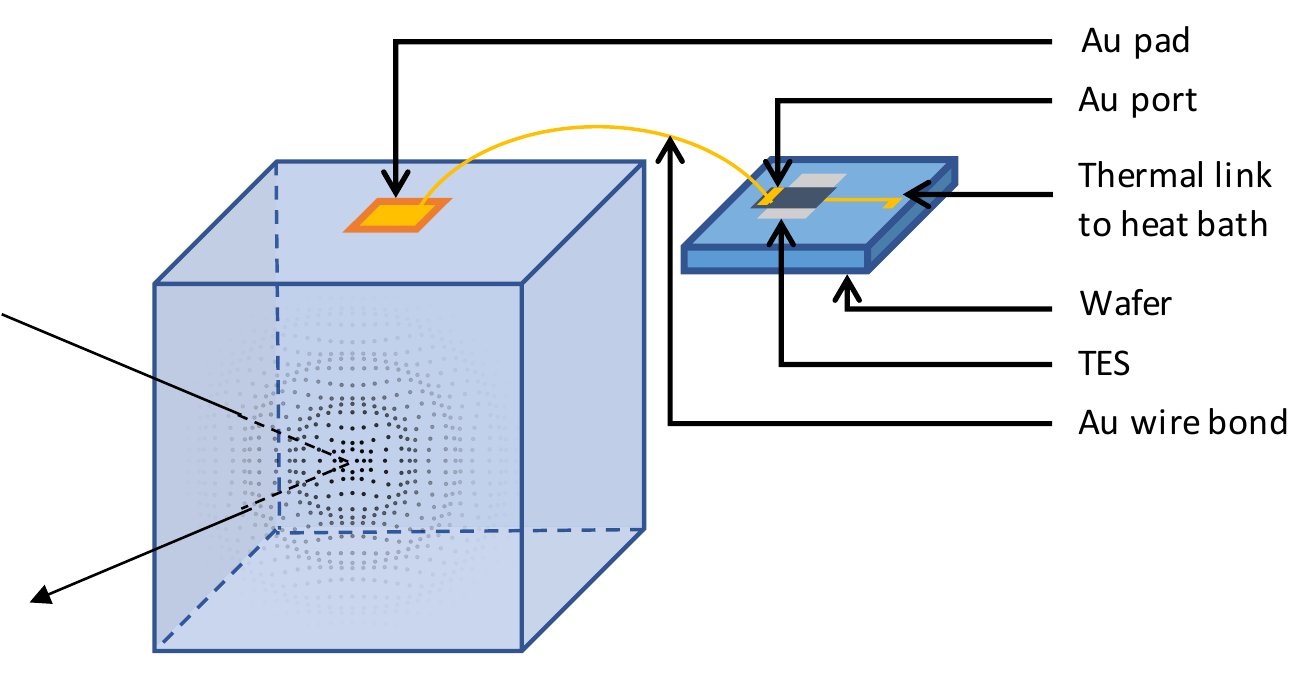}
    \caption{Schematic drawing of the remoTES detector design where the TES is fabricated onto a carrier crystal/wafer. The target crystal is connected to the TES via a Au pad, a Au wire bond, and a Au port.}
    \label{fig:remoTES}
\end{figure}

To operate a NaI crystal -- that is soft and highly hygroscopic -- as a low-temperature detector, COSINUS introduced the so-called remoTES design~\cite{easy_to_fabricate}, where the W-TES is not fabricated directly onto the target crystal, but onto a physically separated wafer. Phonons in the target crystal are collected with a gold (Au) pad located on its surface. The Au pad is coupled to the TES via a Au wire bond as shown in Figure~\ref{fig:remoTES}. This \textit{remote} coupling scheme has two distinct advantages: firstly, the TES and the NaI target crystal are physically separated, thus the crystal can be excluded from the fabrication process of the TES. Secondly, the phonons generated in the target crystal couple directly to the electronic system of the Au pad via the highly efficient electron-phonon coupling.

The remoTES design was first validated in 2021 by using a \unit[(20\x10\x5)]{mm$^3$} silicon (Si) absorber and, in a follow-up study, a \unit[(20\x10\x2)]{mm$^3$} $\alpha$-TeO$_2$ target crystal~\cite{easy_to_fabricate}.  Further studies with Si and NaI confirmed that a single Au wire bond between the target crystal’s phonon collector and the TES provides efficient phonon transmission~\cite{PhD_Kellermann,PhD_Mukund,Master_thesis_KSH}.
Ball bonding on the Au pad is preferred,  as it preserves the soft NaI crystal since it is less destructive than wedge bonding.
Later in 2021, a NaI remoTES was successfully operated both above ground and at a test facility underground at LNGS \cite{cosinus_collaboration_particle_2024,cosinus_collaboration_deep-underground_2024}. 
For the underground measurement, the Au pad was made from a circular gold foil with a radius of~\unit[0.75]{mm}  and a thickness of~\unit[1]{\textmu m}.  The cubic NaI target crystal had a volume of~\unit[1]{cm$^3$} corresponding to a mass of~\unit[3.7]{g}. 

\begin{figure}[t]
    \centering
    \includegraphics[width=\linewidth]{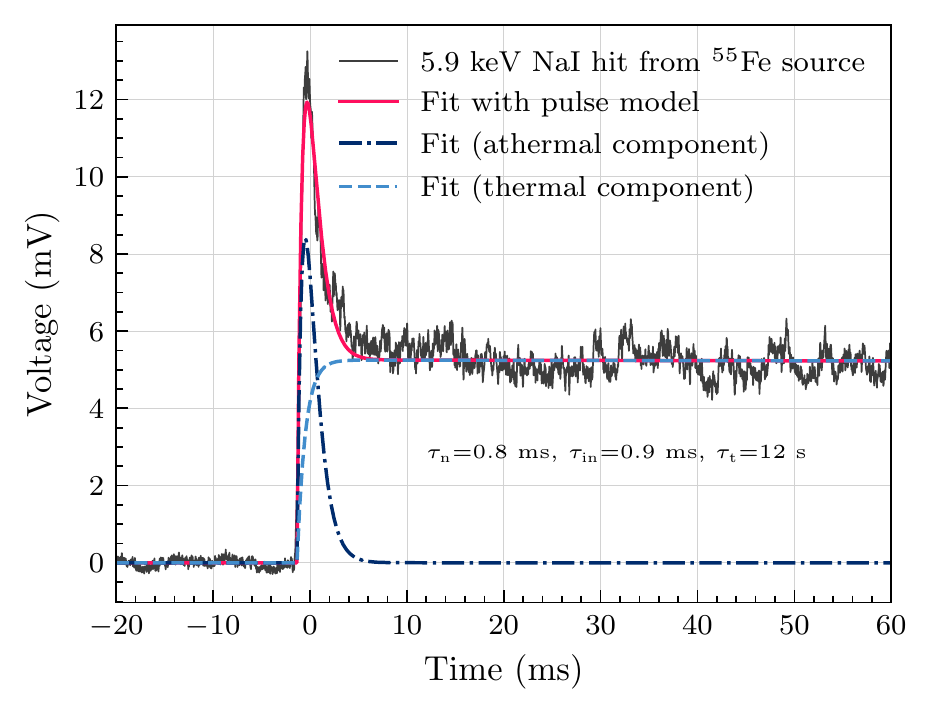}
    \caption{Typical remoTES NaI pulse from an $^{55}$Fe calibration source with a fit to the pulse shape model~\cite{probst_model_1995}. The three time constants governing the pulse shape are shown as well.}
    \label{fig:Frodo_pulse}
\end{figure}

Figure~\ref{fig:Frodo_pulse} shows a typical remoTES-pulse. It demonstrates that the pulse shape model~\cite{probst_model_1995}, developed for directly evaporated TESs as used in CRESST, is also applicable for remoTES pulses. In particular, it can be used to separate the signal into an athermal and a thermal component, which are interpreted as phonons thermalizing in the Au pad and in the NaI crystal, respectively. In NaI, the slowly decaying, thermal component is especially pronounced, which can be ascribed to its high heat capacity.

\begin{figure}[t]
  \includegraphics[width=\linewidth]{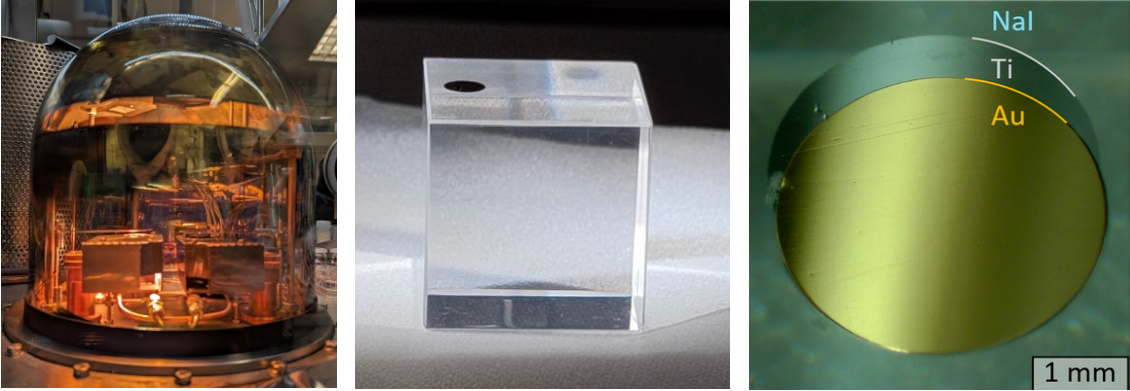}
  \caption{\textit{Left:} An evaporation machine for Au thin film production in operation inside a glove box. \textit{Middle:}~\unit[(21\x21\x21)]{mm$^3$} NaI crystal with Au pad. \textit{Right:} \mbox{1-3\,\textmu m} Au pad evaporated on NaI, including a 10-50\,nm Ti adhesion layer. They are misaligned to illustrate the different thin films~\cite{Master_thesis_KH}.}
\label{fig:evaporator}
\end{figure}

Recently, important steps were taken towards the COSINUS detectors for Run1: depositing the Au port (cf.~Figure~\ref{fig:remoTES}), fully onto the W-film instead of next to it was shown to improve signal coupling by reducing losses into the wafer~\cite{Master_thesis_KSH, ZemaV.2024DaPo}. Furthermore, the Au pad on the NaI crystal can now be evaporated together with a titanium (Ti) adhesion layer instead of the glueing procedure used previously~\cite{cosinus_collaboration_deep-underground_2024,cosinus_collaboration_particle_2024}. Gold evaporation provides pads with the required thickness and strong adhesion, and has the potential of allowing for uniform pad fabrication, which is a benefit for achieving optimal sensitivity. Figure~\ref{fig:evaporator} shows photographs of the evaporator, the crystal used for evaporation, and the resulting Au film. First prototype measurements proved very good performance of evaporated Au pads~\cite{Master_thesis_KH}\footnote{In collaboration with Prof.~A.~Bandarenka, TUM School of Natural Sciences, Physics of Energy Conversion and Storage, Garching, Germany.}.

Finally, the remoTES design presents several benefits, not only for COSINUS but also for cryogenic calorimeters more broadly. By keeping the target crystal separate from TES fabrication, its radiopurity and surface conditions (e.g., oxides: oxygen loss from the lattice in high-vacuum conditions) are better preserved. The design enables reproducible TES production through chip-like batch fabrication on freely chosen substrates. Furthermore, this modular approach may reduce TES-related systematics, allow reuse across different targets, and may facilitate the construction of large detector arrays.

\subsection{Particle discrimination} \label{subsec:PID}
As outlined above, each COSINUS detector module is equipped with a phonon and a light detector measuring the energies $E_\text{p}$ and $L$, respectively.  The ratio of the two is denoted light yield $LY = L / E_\text{p}$. $E_\text{p}$ and $L$ are calibrated using $\upgamma$-sources. For other event types, in particular nuclear recoils, the sharing of the deposited energy between phonon and light is different. Nevertheless, the event-type-independent total deposited energy $E$ can be determined since both quantities are measured~\cite{angloher_likelihood_2024}.

\begin{figure}[t]
    \centering
    \includegraphics[width=\linewidth]{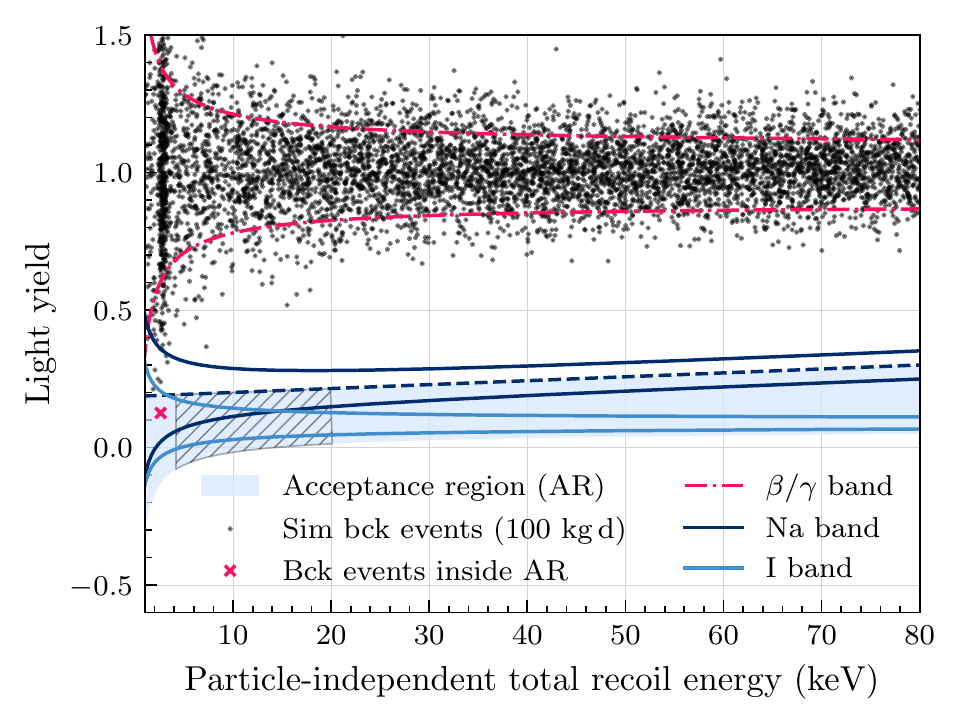}
    \caption{Particle discrimination on an event-by-event basis allowing for a quasi-background-free experiment. Background events from a simulation of 100\,kg\,d only rarely leak into the acceptance region (defined between the mean light yield of the Na band to the 99\,\% lower limit of the I band). For a measurement demonstrating the discrimination power, see Figure~\ref{fig:summer_band}. The hatched area marks the region of exceptional sensitivity to the DAMA signal, cf.~Figure~\ref{fig:leakage_sensitivity_discovery_potential}.}
    \label{fig:bands_discrimination}
\end{figure}

Figure~\ref{fig:bands_discrimination} shows simulated data in the light yield ($LY$) versus energy ($E$) plane for an assumed exposure of 100\,kg\,d before cuts. The distribution of events in this plane can be accurately modeled, which is presented in detail in~\cite{angloher_likelihood_2024}; here, only the most relevant aspects are mentioned. The simulation follows projections based on previous prototype measurements~\cite{cosinus_collaboration_deep-underground_2024,cosinus_collaboration_particle_2024}, as detailed on page~\pageref{sec:projections}.

Electrons and $\upgamma$s interact with the electrons of the target material and produce the highest light output. Therefore, their mean light yield is set to one by calibration. The spread of the light distribution around this mean (called a band) is modeled by a Gaussian function whose energy-dependent width is dominated by the noise of the light detector and the Poissonian nature of light production. In Figure~\ref{fig:bands_discrimination}, the solid lines correspond to the 90\,\% upper and lower boundaries of the bands, thus 80\,\% of the respective event class are expected inside the boundaries. 

Nuclear recoils are quenched (emit less light) and appear in two separate bands in Figure~\ref{fig:bands_discrimination}, dark blue for recoils off Na and light blue for recoils off I. The positions and slopes of the bands are determined by the energy-dependent quenching factors (QFs).

Figure~\ref{fig:bands_discrimination} demonstrates the excellent discrimination power of a COSINUS detector module: for the assumed background level and target exposure of 100\,kg\,d, only one event (magenta cross) is expected to leak into the so-called \emph{acceptance region} (shaded blue) that includes all events below the mean of the Na band (dashed line) and above the lower 99\,\% lower boundary of the I band. This choice constitutes a good compromise between minimizing electro-magnetic background leakage and maximizing signal expectation in the nuclear recoil bands. The hatched area marks the region with high event expectation for a DAMA-compatible DM signal at practically zero background.

\begin{figure}[t]
    \centering
    \includegraphics[width=\linewidth]
    {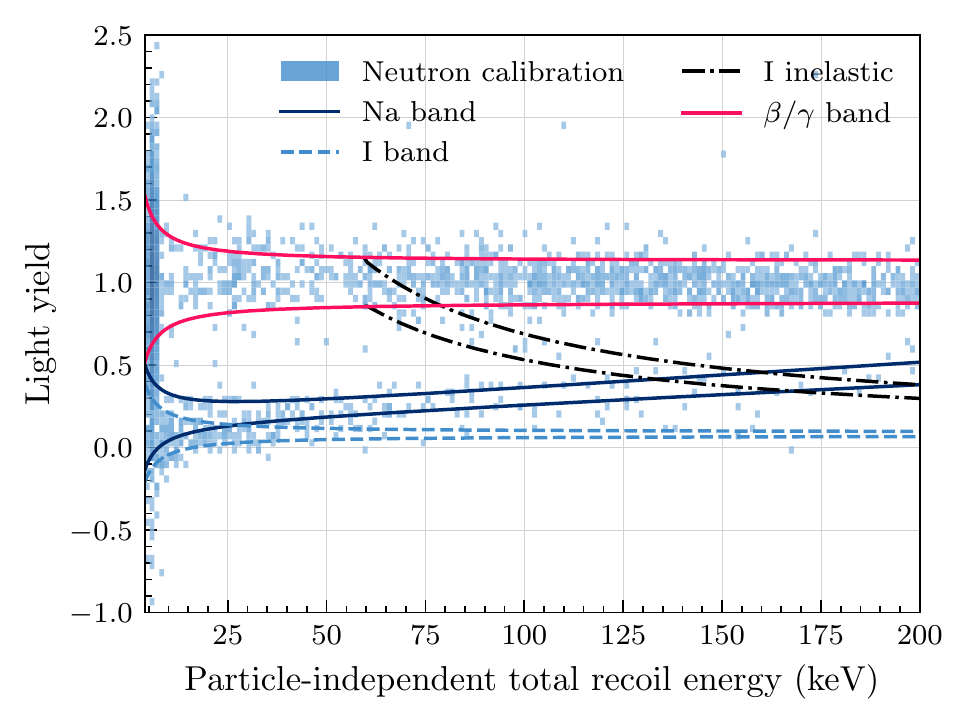}
    \caption{Data from \cite{cosinus_collaboration_deep-underground_2024} with fitted band description for the \ega~background and nuclear recoil bands due to neutrons from an AmBe source. The large number of low-energy events in the \ega~band at an energy of $\sim$6\,keV is caused by an $^{55}$Fe source which will be absent in the final experiment.}
    \label{fig:summer_band}
\end{figure}

The band description was successfully validated in prototype measurements: In~\cite{cosinus_collaboration_particle_2024}, a NaI crystal read out with a remoTES and operated together with a silicon-on-sapphire wafer light detector achieved a baseline resolution of ($2.07\pm0.01$)\,keV in event-type-independent energy $E$. This above-ground measurement proved particle identification on an event-by-event basis in a COSINUS NaI detector: nuclear recoil events from an AmBe neutron calibration source could be clearly distinguished from electronic recoils down to the analysis threshold of 20\,keV. A second NaI prototype with a mass of 3.7\,g was operated underground at LNGS in a test facility of the CRESST collaboration. It featured an improved baseline resolution of ($441\pm11$)\,eV~\cite{cosinus_collaboration_deep-underground_2024}. Figure~\ref{fig:summer_band} shows its neutron calibration data in the light yield versus energy plane, including the fitted bands for electronic recoils, nuclear recoils off Na and I, and inelastic n-scatters off I. This measurement allowed to determine the nuclear QFs and their energy dependence in-situ by treating them as free parameters in the maximum likelihood band fit~\cite{angloher_likelihood_2024}. 
The simulation in Figure~\ref{fig:bands_discrimination} is based on those results, yet all QFs will also be determined in-situ and individually for each detector module in the final experiment. A direct, event-by-event particle identification and in-situ measurement of the quenching factors present a distinct advantage of operating NaI as a scintillating calorimeter.

Two things are noteworthy for NaI when compared to other scintillators as in particular CaWO$_4$ for CRESST: firstly, no significant difference in light production between electrons and $\upgamma$s was observed \cite{lang_scintillator_2009,angloher_likelihood_2024}, thus, a single band can be used (magenta in Figure~\ref{fig:bands_discrimination}). Secondly, the \ega-band does not bend down towards low energies, an effect known as non-proportionality \cite{angloher_likelihood_2024,lang_scintillator_2009}.

Despite the limited exposure of only 11.6\,g\,d in this measurement, first COSINUS DM results were extracted, shown in green lines in Figure~\ref{fig:limitplot}. Both lines correspond to 90\,\%\,C.L.~exclusion limits on the DM-nucleon cross-section calculated using the Yellin optimum interval method \cite{yellin_finding_2002,yellin_extending_2007,noauthor_optimum_2011}. Dashed and solid lines refer to exclusion limits obtained by ignoring or including the particle discrimination power of the detector, respectively. Their comparison demonstrates that particle discrimination allowed about two orders of sensitivity improvement, down to DM particle masses of $m_{\chi}\sim 10~$GeV, where the effect of events from the (optional) $^{55}$Fe calibration source leaking into the region of interest is visible\footnote{Such internal sources will not be used in the final measurement.}. 

\subsection{\texorpdfstring{\onepi}~~Run1 detector design} \label{subsec:1piDesign}
\begin{figure}[t]
    \centering
    \includegraphics[width=\linewidth]{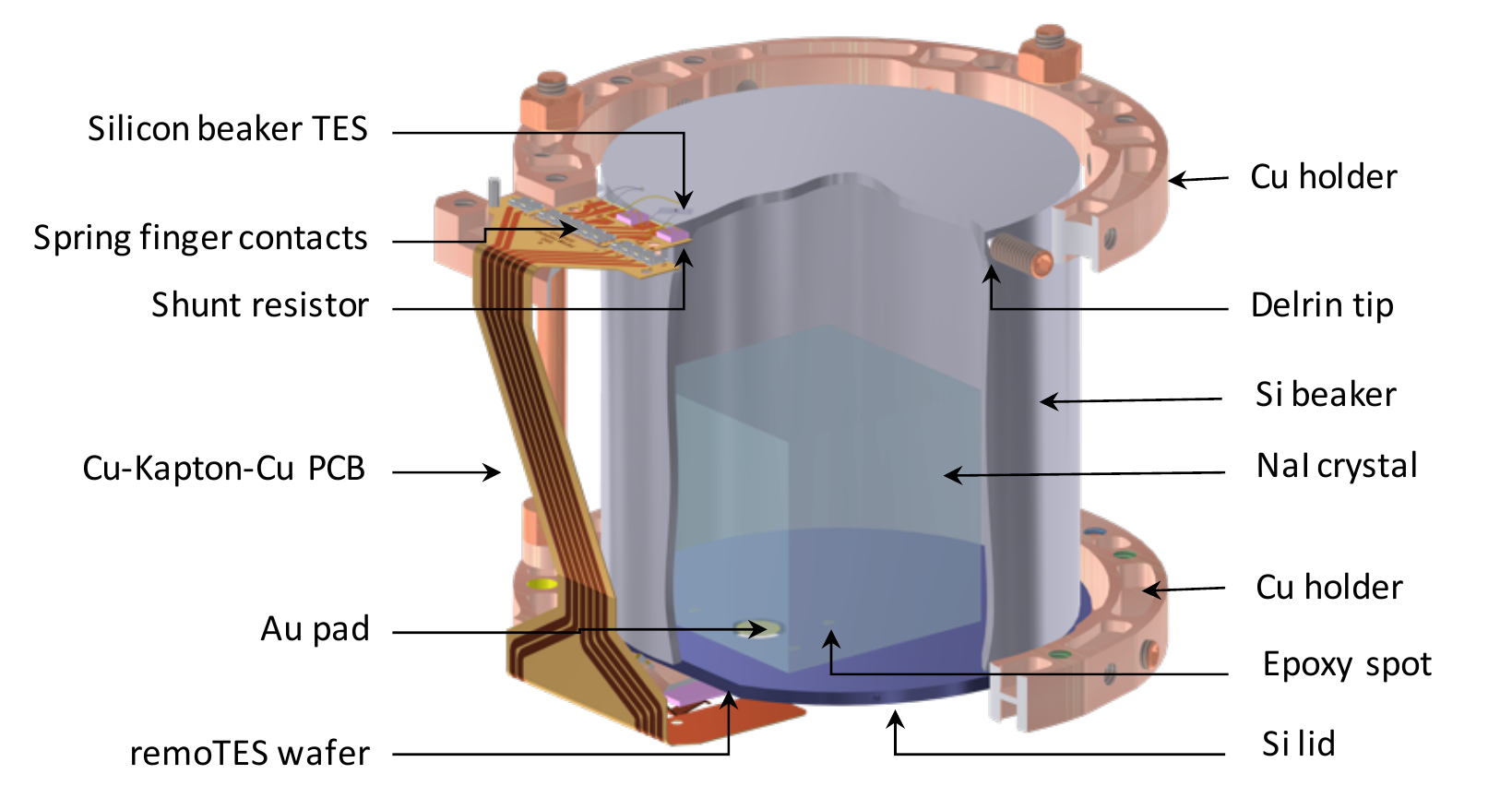}
    \caption{Sketch of a detector module for \onepi~Run1 with a 34.8\,g cubic NaI target crystal.
    }
    \label{fig:Modul}
\end{figure}

Details of a COSINUS Run1 detector module using a \unit[(21\x21\x21)]{mm$^3$} NaI target crystal, designed without compromise to optimize light detection efficiency while maintaining stringent low-background conditions, are depicted in Figure~\ref{fig:Modul}. 

The radiopure, undoped NaI crystals for COSINUS were produced at SICCAS (Shanghai Institute for Ceramics, Chinese Academy of Sciences) in Shanghai, China, using ultrapure Astrograde powder from Merck. To date, a total of 28 cubic crystals of \unit[(21\x21\x21)]{mm$^3$} with a mass of 34.8\,g were produced from five ingots. All crystals were optically polished on each face and sealed before being delivered by boat to LNGS. The internal radioactive contamination level was studied by screening (HPGe and ICP-MS) and revealed 6-22\,ppb of $^{40}$K (4\,ppb in the initial powder) and $<$1\,ppb for both U and Th\footnote{ICP-MS measurements performed at the INFN LNGS chemistry department, publication in preparation.}, satisfying the required radiopurity goal of being on the radiopurity level of DAMA/LIBRA. 

The NaI crystal is glued to a silicon lid (\unit[40]{mm} diameter and \unit[1]{mm} thickness) with the gold pad aligned to a hole for the Au wire bond. Gluing is done with a machine to make five reproducible dots. The TES is fabricated onto a separate sapphire wafer with a Au port design (see Figure~\ref{fig:remoTES}). A single Au ball bond connects the absorber to the TES.
Since the NaI target crystals are highly hygroscopic, assembly of detector parts and detector mounting have to be exclusively carried out in a controlled atmosphere by utilizing a dedicated glove box with a low humidity level $<$100 ppm.

A silicon beaker (\unit[41]{mm} diameter and height, \unit[15.4]{g}) and the silicon lid enclose the NaI crystal, serving as both effective scintillation light absorber and as a veto against potential surface \textalpha-particle backgrounds\footnote{If the silicon lid is also instrumented this provides an active 4\textpi-veto.} \cite{angloher_results_2012,kuzniak_surface_2012}.
As a key design change, adaptable shunt resistors are now directly integrated into each module, replacing the previously shared external design and enabling individual detector readout optimization.

\section{COSINUS setup at the LNGS underground laboratories} \label{sec:setup}
\begin{figure}[t]
    \centering 
    \includegraphics[width=\linewidth]{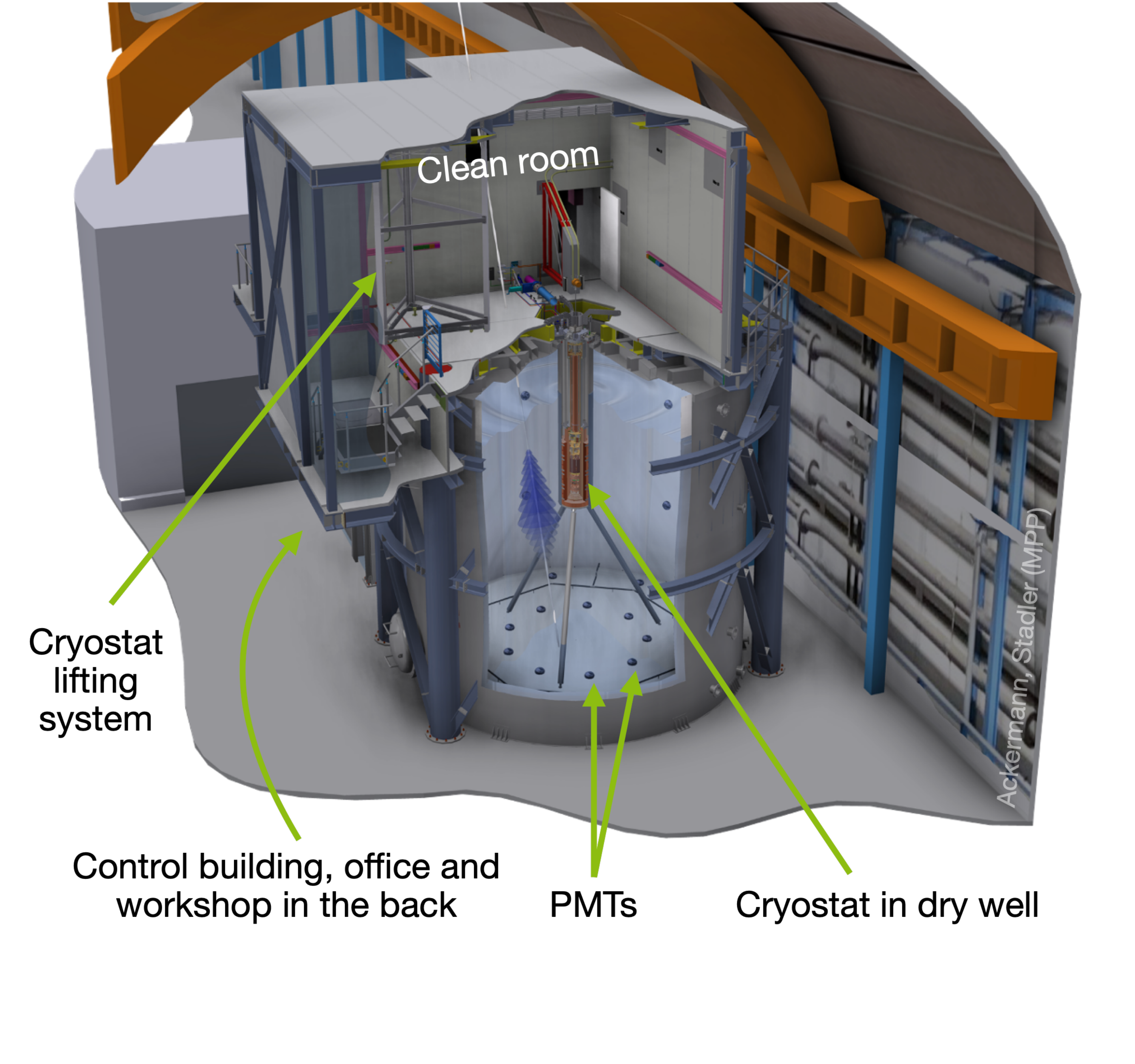}
    \caption{Rendering of the COSINUS setup in hall B of LNGS with the cryostat in the center of the cylindrical water tank. Depicted is the Cherenkov light emitted by a particle traversing the water tank, which is picked up by 30 PMTs. A clean room is installed above the water tank to allow the handling of the cryostat (and detectors) in a clean environment.}
    \label{fig:facility}
\end{figure}

The COSINUS experimental underground facility was designed  following~\cite{angloher2021simulationbased} and consists of four main parts (see Figure~\ref{fig:facility}): i) the dry dilution refrigerator supplies the low temperatures required to operate the detectors at mK temperatures, ii) the water tank operated as active Cherenkov-muon veto ensures low radiogenic and cosmogenic neutron background conditions at the detector stage, iii) a clean room allows to mount detectors and to move the cryostat in and out of the dry well with the help of a lifting system and iv) a three-story control building next to the water tank.

The step-by-step installation and commissioning of the main components of the experimental facility at LNGS -- including the water tank, servicing level with cleanroom and control building, as well as the electrical infrastructure -- began in 2021 and was completed in 2023. 

\subsection{Cryostat and data acquistion}
Installation of the custom dry dilution refrigerator from CryoConcept\footnote{An affiliate of Air Liquide and shut down since 28.03.2025.} at LNGS commenced in spring 2024. During the first cooldown, the cryostat successfully reached a base temperature of 9.3\,mK and demonstrated the required cooling power.

In a second commissioning run in July 2024, the system was equipped with internal passive shielding: a 190\,kg block of ultra-pure copper mounted on the Still plate at approximately \unit[800]{mK}. This cooldown was also successful, reaching base temperature within 180\,h -- well within expectation. Since then, two additional cooldowns have been carried out to assess long-term operational stability, evaluate vibration levels, and validate the detector plate decoupling system. 
To avoid inductive effects and noise from microphonic, piezoelectric, or triboelectric sources, the cables will be solidly and securely fixed to the refrigerator. A custom-designed decoupling system, consisting of three bronze springs with a resonance frequency of 2\,Hz and anchored to a 190\,kg copper block, provides effective vibration damping and enables stable detector operation~\cite{PhD_Kellermann,kellermann_vibration_2024}.

COSINUS developed a modular custom-designed cabling solution for a total of 48 TES channels, installed in summer 2025. 

For detector operation and signal acquisition, COSINUS uses an in-house developed integrated system, denoted Versatile Data Acquisition (VDAQ). For each TES channel, the VDAQ has two DACs for TES bias and heater current, respectively, and one high-resolution ADC (24\,bit) for continuous, dead-time free signal acquisition from the SQUID\footnote{SQ300 DC-SQUIDs from STARCRYO, \href{https://starcryo.com}{starcryo.com}}. The VDAQ periodically injects electrical pulses to the heater to map the detector response over the full dynamic range of the TES and as a function of time.

The VDAQ is jointly developed for several experiments (COSINUS, CRAB \cite{abele_crab_2025}, CRESST,  NUCLEUS \cite{strauss_$nu$-cleus_2017}) and was already successfully used for data taking~\cite{abele_crab_2025}.

\subsection{Muon veto} \label{subsec:muonveto}
Despite the 3600\,m water equivalent rock overburden of LNGS, which reduces the muon flux by a factor of $10^6$ \cite{ambrosio_vertical_1995}, muon-induced neutrons remain a dangerous background: previous simulations predict a rate of ($3.5\pm0.7$)\,counts\,/\,kg\,\,year ($\sim10$\,counts\,/\,1000\,kg\,d) for cosmogenic neutrons hitting the detector \cite{Angloher2024}.  Reducing this number requires an active muon veto, which is particularly challenging the slow cryogenic COSINUS detectors with a time resolution of $\mathcal{O}$(10\,ms). A high trigger rate of the veto system would result in a high number of accidental coincidences between the veto and cryogenic detectors and, therefore, cause large dead times. The design of the COSINUS muon veto is based on Monte Carlo simulations, all details may be found in \cite{Angloher2024}.

The COSINUS muon veto is a classic  water Cherenkov detector measuring charged particles via their emitted Cherenkov light when traversing ultra-pure water. It consists of 30 photomultiplier tubes (PMTs)\footnote{\href{https://www.hamamatsu.com/jp/en/product/optical-sensors/pmt/pmt_tube-alone/head-on-type/R5912.html}{Hamamatsu R5912-100}} in the COSINUS water tank (7\,m height and diameter) surrounding the detectors, see Figure~\ref{fig:facility}. 18 PMTs are placed on the bottom of the tank and 12 on its side walls.

\begin{figure}[t]
    \centering
    \includegraphics[width=\linewidth]{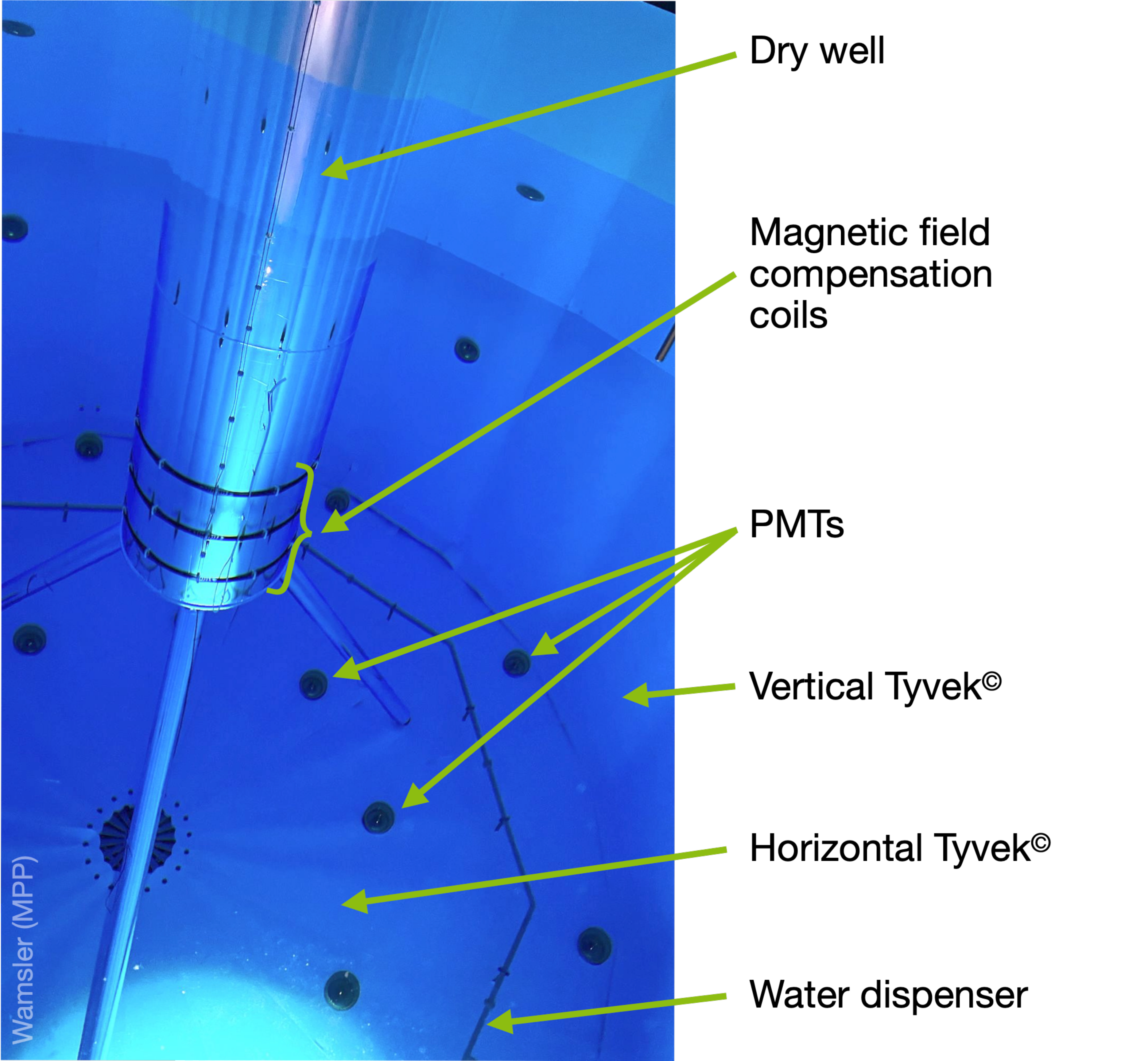}
    \caption{Photograph taken during the filling of the water tank from the top which took place in Jan/Feb 2025.}
    \label{fig:watertankphoto}
\end{figure}

COSINUS optically divided the water tank into an inner/active part and an outer/non-active (dead) part using reflecting Tyvek\textsuperscript{\textcopyright} tri-layer curtains mounted at a distance of 30\,cm from the tank walls and the tank floor, see Figure~\ref{fig:watertankphoto}. They prevent \gas~originating from the steel tank walls or the outside from reaching the inner/active part of the muon veto and, therefore, reducing the overall (false) trigger rate of the system. We find that the rate is halved for every 10\,cm of dead layer for thicknesses smaller than 30-40\,cm~\cite{Angloher2024}. 

Our simulations indicate that the proposed muon veto layout, combined with a  5-fold PMT coincidence trigger requirement, results in a manageable overall (false) trigger rate of less than 1\,Hz while maintaining a total veto efficiency of 97\,\% (99.6\,\% for muon events and 44.4\,\% for shower events, respectively). This efficiency corresponds to a muon-induced neutron rate of ($0.11\pm0.02$)\,counts\,/\,(kg\,y) or $\sim 0.3$\,counts\,/\,(1000\,kg\,d). Thus, even for the target exposure of Run2 the expectation is less than one muon-induced neutron event.

Although primarily designed as a muon/neutron shield, the veto system can also be used, as is, to effectively detect neutrinos from core-collapse supernovae~\cite{angloher_neutrino_2025}. Hence, COSINUS could contribute to the Supernova Early Warning System~\cite{AlKharusi2021} in the future. 

\section{Projections} \label{sec:projections}
In this section, projections regarding the achievable sensitivity of COSINUS are presented. The calculations are based on the measurement of the 3.7\,g prototype~\cite{cosinus_collaboration_deep-underground_2024} and use an electrothermal feedback model~\cite{Wagner2023} to predict the achievable performance of the 34.8\,g detectors which will be used in \onepi~Run1. The model predicts that a (nuclear recoil) energy threshold for the phonon detector of 0.68\,keV may be achieved by simple adaptations of the prototype -- a lower transition temperature $T_c$ of the TES together with an optimized gold pad size -- in combination with the operation in a low background environment. For the projections shown here, a resolution of $\sigma=0.2$\,keV is conservatively assumed, resulting in a threshold of $\ethr=1$~keV, which is also compatible with~\cite{Angloher2016}. While the phonon performance discussed above determines the energy threshold of the detector, the light channel performance dictates the discrimination power between electronic and nuclear recoils. A light detector resolution of $\sigma_{\mathrm L}=0.11$~keV$_{\mathrm{ee}}$ (electron-equivalent energy) is assumed~\cite{angloher_cosinus_2019}. 
Assumptions on the background are compatible to DAMA: a constant electron background of 1~count\,/\,(keV$_\text{ee}$\,kg\,d) plus an intrinsic $^{40}$K activity of 600~\textmu Bq\,/\,kg. Monte Carlo simulations of external and internal neutron backgrounds revealed a total expected rate of only 0.156~counts\,/\,(kg\,y)~\cite{Angloher2024, Fuss2022} which is therefore neglected in the following. The survival probability of events is estimated to be 50\,\% at threshold, and to smoothly increase until it stays at a constant value of 76\,\%\footnote{This value assumes 10\,\% signal loss due to detector instabilities, 10\,\% loss due to quality cuts, as well as 4\,\% due to random coincidences with the muon veto (corresponding to a 1\,Hz muon veto dark count rate and a $\pm$\,20\,ms coincidence window).} above 3~keV\footnote{I.e.~the efficiency used is given by $\epsilon(E)=0.5+0.26~\mathrm{Erf}(E-\ethr)$.}.

\begin{figure*}[t]
    \centering
    \includegraphics[width=\linewidth]{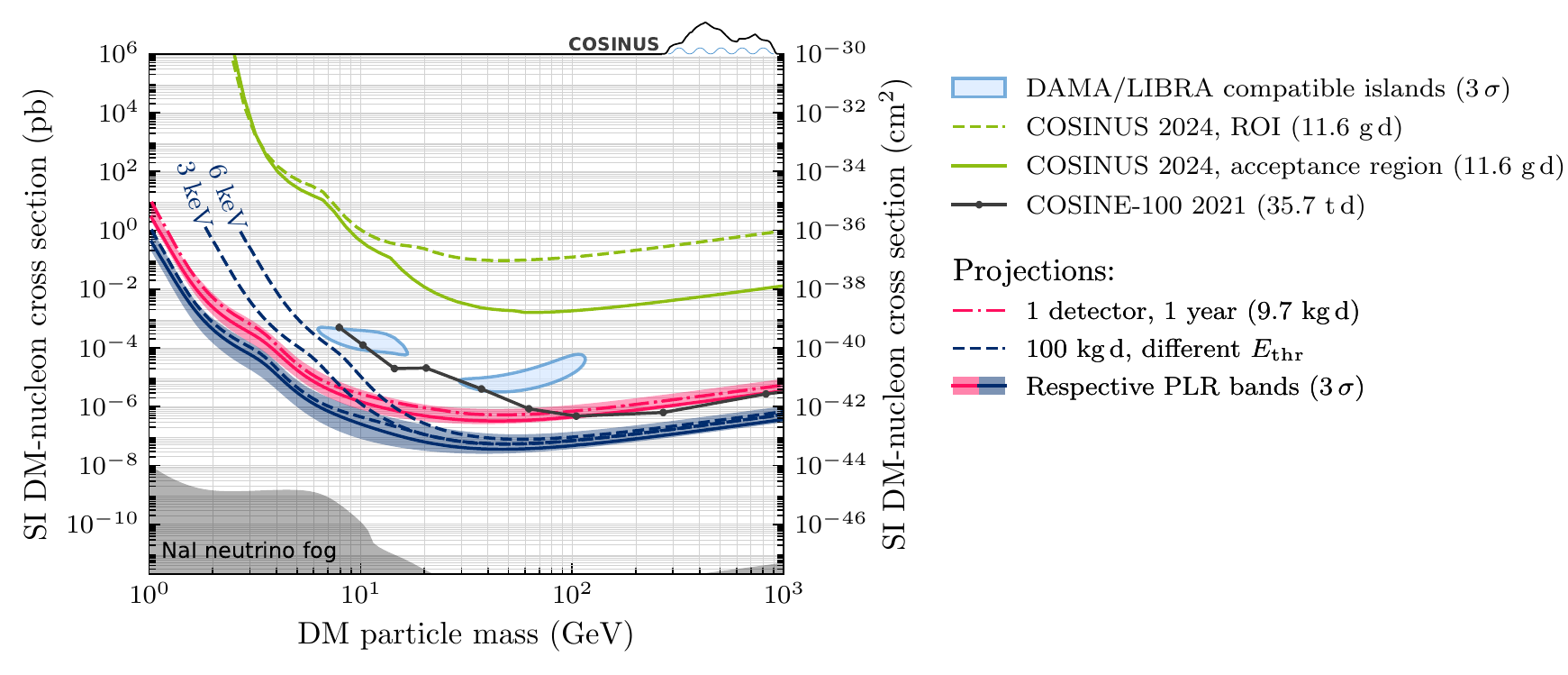}
    \caption{Projections of the COSINUS exclusion power in the standard scenario. The dashed lines show standard Yellin limits, the solid lines PLR median and $3\,\sigma$ confidence region. Once the nominal exposure of 100\,kg\,d is achieved, a threshold as high as 6~keV is sufficient to rule out the DAMA/LIBRA islands (taken from \cite{savage_compatibility_2009}) in the standard scenario. A total of 8 detector modules is anticipated to collect this exposure in the first phase of COSINUS. However, a single module with a threshold of 1~keV has enough exclusion power after only one year of data-taking. The green lines show limits previously obtained in a COSINUS prototype measurement~\cite{cosinus_collaboration_deep-underground_2024}. Neutrino fog data from \cite{PhysRevLett.127.251802}, COSINE-100 limit from \cite{adhikari_strong_2020}.}
    \label{fig:limitplot}
\end{figure*}

\subsection{Sensitivity in the standard scenario}
First, the sensitivity in the so-called standard scenario is presented, with standard assumptions on the dark matter velocity spectrum (Maxwellian) and for elastic DM nucleus scattering \cite{baxter_recommended_2021}. Figure~\ref{fig:limitplot} shows the result, together with the interpretation of DAMA in that case~\cite{savage_compatibility_2009}. Magenta lines and bands correspond to a gross exposure of 9.7\,kg\,d, which can be achieved with a single 34.8\,g detector running for one year. Very similar sensitivities are reached whether Yellin's optimum interval method~\cite{yellin_finding_2002,yellin_extending_2007,noauthor_optimum_2011} is applied to events in the acceptance region (dashed line) or a profile likelihood method~\cite{angloher_likelihood_2024} based on all data and the previously described band description is used (solid line and confidence belt). As can be seen, a single detector module would suffice to clarify the standard scenario. The sensitivity is further improved by increasing the exposure to 100\,kg\,d (dark blue lines and belts), as collected by all eight modules within about one and a half years. With 100\,kg\,d of exposure a clarification of the DAMA signal in the standard scenario would even be achievable if the threshold was worse than expected (1\,keV), see the dark blue dashed lines marked for thresholds of 3\,keV and 6\,keV, respectively.

\subsection{Model-independent DAMA/LIBRA cross-check}
Although COSINUS measures a signal-only rate of nuclear recoils and DAMA/LIBRA measures a modulating event rate over an unknown background, a model-independent comparison is straightforward due to the fact that both experiments use NaI as a target material. 

The most conservative approach is based on the fact that the modulation amplitude measured in DAMA~\cite{bernabei_first_2018,bernabei_further_2021,bernabei_annual_2025} cannot be larger than the average rate measured in COSINUS which is presented in detail in~\cite{kahlhoefer_model-independent_2018}. The finding is that roughly 1000\,kg\,d of exposure are needed to exclude a DM origin of the DAMA/LIBRA signal with no assumptions on the DM interactions beyond the requirement of producing nuclear recoils. 

\begin{figure*}[t]
    \centering
    \includegraphics[width=\linewidth]{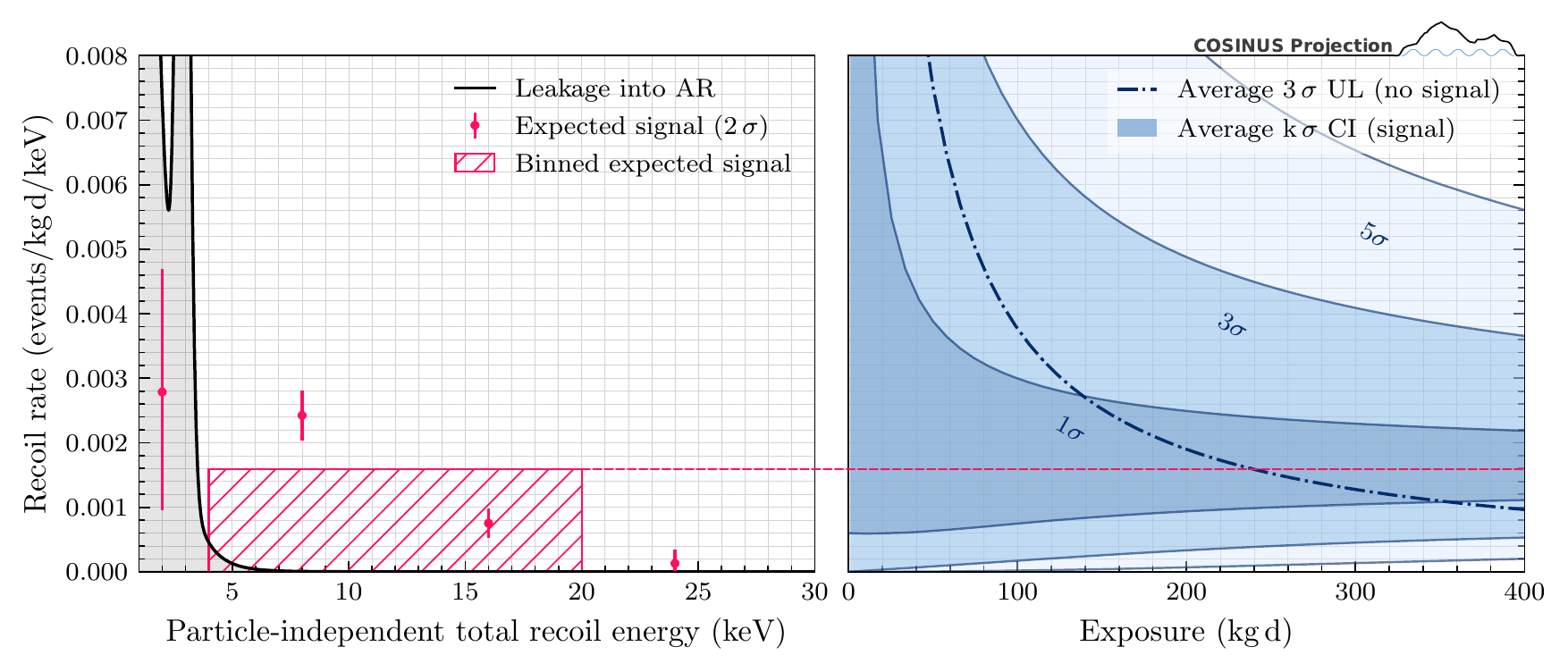}
    \caption{Discrimination power and sensitivity of COSINUS. Left: Expected event rate from background processes into the AR (blue area in Figure~\ref{fig:bands_discrimination}) as well as the expected signal rates obtained from unfolding the DAMA spectrum~\cite{angloher_cosinus_2025}. The signal is collected in the [4, 20] keV bin (corresponds to hatched region in Figure~\ref{fig:bands_discrimination}) for a sensitivity projection. Right: Average upper limit that can be placed if no signal is present in the hatched bin, as well as resulting confidence interval on the rate if a signal \textit{is} present, as a function of collected exposure. Confidence intervals are calculated according to Feldman\,\&\,Cousins~\cite{feldman_unified_1998}.}
    \label{fig:leakage_sensitivity_discovery_potential}
\end{figure*}

However, the above approach neglects the information contained in the energy-dependence of the modulation amplitude as measured by DAMA/LIBRA because it considers one energy bin only. In \cite{angloher_cosinus_2025} the entire DAMA/LIBRA modulation spectrum is unfolded, and a lower-bound prediction of the energy spectrum in COSINUS detectors in presence of the DAMA signal is obtained, using the same argument but now \emph{for each bin}. It is depicted with magenta data points in Figure~\ref{fig:leakage_sensitivity_discovery_potential} (left), together with the expected leakage from \ega-backgrounds into the acceptance region (cf. Figure~\ref{fig:bands_discrimination}). Here, the hatched magenta region can be identified to be particularly suitable for comparing the potential signal to the expected background leakage in a COSINUS detector. The height of the hatched bin represents the average signal rate in COSINUS between 4 and 20\,keV. The right side of Figure~\ref{fig:leakage_sensitivity_discovery_potential} shows the significance for a positive confirmation (blue belts) in presence, as well as a 3\,$\upsigma$-exclusion (dashed-dotted line) in the absence of the signal, as a function of collected exposure. Using the rate predicted in the magenta hatched area on the left, one achieves a 3\,$\upsigma$-exclusion with an exposure of about 250\,kg\,d. Conversely, a 5\,$\upsigma$-positive signal confirmation could require as little as 150\,kg\,d. The study of \cite{angloher_cosinus_2025} clearly shows that incorporating the information from the full DAMA energy spectrum significantly relaxes the requirements on exposure and background in the acceptance region. For more details, including a demonstration of the robustness of the result for different unfolding approaches, the reader is referred to \cite{angloher_cosinus_2025}.

\section{COSINUS physics program} \label{sec:PhysicsProgram}
The primary motivation and goal of COSINUS is to cross-check the DAMA/LIBRA signal. At the same time, its modern and unique low-background cryogenic facility allows for many more physics cases to be pursued in the future. This section both outlines the road map for the DAMA/LIBRA cross-check and highlights some of those additional projects.

\subsection{NaI program} \label{subsec:NaIProgram}
COSINUS adopts a staged approach to achieve its primary goal of scrutinizing the long-standing dark matter (DM) claim reported by the DAMA/LIBRA collaboration. The initial phase will involve eight NaI detector modules, each with a mass of 34.8\,g, to collect a DM exposure of 100\,kg\,d over a period of 12-18 months, including times for detector setup and calibration. An exposure of 100\,kg\,d is sufficient to confirm or refute the DAMA signal under standard assumptions (see Figure~\ref{fig:limitplot}). It will also allow us to place the first constraints in a model-independent cross-check framework (see Figure~\ref{fig:leakage_sensitivity_discovery_potential}). To fully pursue this model-independent investigation, significantly higher exposures are required. Therefore, in \onepi~Run2, COSINUS plans to collect 1000\,kg\,d using 20 detector modules, each with a 100\,g NaI crystal, running for approximately two years. Two full annual cycles will also provide a decent level of sensitivity to include time information as an additional observable to further enhance signal-to-background discrimination. 

In the case of a positive signal observed in \onepi, a possible next phase might follow, \twopi, dedicated to investigating the annual modulation as a solid signature for a galactic DM origin of a potential signal. The profit, feasibility, and design of \twopi~also depend on the broader experimental landscape and the results emerging from other ongoing efforts in the field.

\subsection{Additional new physics cases}
In addition to its primary NaI-based DM search program, COSINUS is actively involved in the development of advanced cryogenic detector technologies geared towards future rare event searches: 

i) Achieving reliable and consistent TES fabrication remains a key challenge. To address this, a partnership with the MPG Halbleiterlabor (HLL) was established to scale production to a semi-industrial level, enabling controlled processes, standardized protocols, and regular quality checks -- essential for next-generation rare event searches with large arrays of cryogenic detectors.

Large TES sensor arrays demand a transition from single-channel SQUID readout to multiplexed readout. To support next-generation searches, new analysis strategies are advancing in parallel with detector development, including the statistically robust combination of data from numerous channels and automating detector operating conditions and data processing. 

ii) The remoTES readout decouples the sensor from the absorber, enabling the use of a broader range of materials. We are testing hydrogen-rich crystals for low-mass DM sensitivity, and superconductors for their potential in DM–electron scattering and reduced thermal stress, possibly mitigating the low-energy excess (LEE) \cite{baxter_low-energy_2025,adari_excess_2022}. Together with our partners of the envisioned CRYOCLUSTER, we are also exploring SiO$_2$ as a thermally compatible absorber with W-TESs to help suppress the LEE.

iii) The O\textnu DES project aims to use NaI or GaAs scintillators for the detection of single photons from DM-electron scattering using an array of cryogenic photon detectors surrounding the target crystal~\cite{Zema2024a}. 

iv) The DAREDEVIL project investigates the use of GaAs as low-threshold cryogenic detectors with NTD readout to study DM-electron interaction scenarios \cite{Helis:2024vhr,Melchiorre_2025}. 

\section{Conclusion}

COSINUS is the only NaI experiment operating NaI crystals as low-temperature detectors. While \textit{standard} room-temperature NaI experiments only measure scintillation light, a low-temperature detector provides a dual-channel readout of phonon and light which significantly improves sensitivity at low (nuclear recoil) energies, thanks to the low energy threshold of the phonon channel. Furthermore, the dual readout provides powerful background rejection on event-by-event basis. COSINUS is currently being commissioned at the LNGS underground laboratories. The first data-taking campaign, \onepi~ Run1, involving the operation of eight NaI modules for 12-18 months, is scheduled to begin in late 2025. With the target exposure of 100\,kg\,d, COSINUS will be able to completely rule out or confirm a DM origin of the DAMA/LIBRA signal in the standard scenario and will start to severely constrain any DM-nucleus scattering origin in a model-independent way. In \onepi~Run2, COSINUS will increase the total NaI target mass to collect a maximum exposure of 1000\,kg\,d. Run2 will deliver an unambiguous test of whether the DAMA/LIBRA modulation signal arises from DM-nucleus scattering, independent of assumptions about the dark matter halo and/or interaction mechanism.

\subsection{Outlook}
The COSINUS facility was designed and constructed respecting the needs of an ultra-low background budget while providing about 42 liters of experimental volume at $\unit[<10]{mK}$. This makes COSINUS an ideal place for future next-generation rare event searches. Thus, in parallel to its NaI-based dark matter search, COSINUS is already working towards next-generation cryogenic detectors also in the frame of the envisioned CRYOCLUSTER aiming to bring together TES-based rare event searches in Europe. This includes scaling TES production to semi-industrial standards at HLL, introducing multiplexed readout and automation, low-energy-excess mitigation studies, and exploring remoTES designs with alternative target materials for future rare event searches.

\section*{Acknowledgments}
We are also grateful to LNGS for their generous support and persistent technical and administrative assistance. This work was supported by the Austrian Science Fund (FWF), projects ALCDA \url{http://dx.doi.org/10.55776/PAT1239524} and Scies4Free \url{http://dx.doi.org/10.55776/DFH66}, the Research Council of Finland (grant\# 342777), and the Klaus Tschira Foundation.\\

\noindent\textbf{Author contributions}\par
All authors contributed to the research presented in this paper and have read and agreed to this version of the manuscript. Authors are listed alphabetically by their last names.
Manuscript preparation: F.~Reindl, K.~Sch\"affner, P.~Schreiner.\\

\noindent\textbf{Competing interests}\par
The authors declare to have no competing interests.\\

\noindent\textbf{Data availability}\par
This manuscript has no associated data.\\

\noindent\textbf{Materials \& Correspondence}\par
Correspondence and material requests should be addressed to the corresponding authors or to \href{mailto:naice@cosinus.it}{\texttt{naice@cosinus.it}}.

\section*{References}
\bibliographystyle{h-physrevFR}
\bibliography{literature,NatureCOSINUSFlo}

\end{document}